\documentclass{aa}
\usepackage{times}
\usepackage{graphics}
\begin{document}
\topmargin0.0cm
\thesaurus{01(03.20.1; 08.03.1; 08.03.4; 08.09.2: IRC\,+10\,216; 08.13.2; 08.16.4)}
\title{76\,mas speckle--masking interferometry of IRC\,+10\,216 with the SAO 6\,m
telescope: Evidence for a clumpy shell structure}
\author{
G. Weigelt\inst{1}\and
Y. Balega\inst{2}\and
T. Bl\"ocker\inst{1}\and 
A.J. Fleischer\inst{3}\and
R. Osterbart\inst{1}\and
J.M. Winters\inst{3}
}
\institute{
Max--Planck--Institut f\"ur Radioastronomie, Auf dem H\"ugel 69,
D--53121 Bonn, Germany
\and
Special Astrophysical Observatory, Nizhnij Arkhyz, Zelenchuk region, 
Karachai--Cherkesia, 35147, Russia
\and
Technische Universit\"at Berlin, Institut f\"ur Astronomie und
Astrophysik, Sekr.  PN 8-1, Hardenbergstr.  36, D--10623 Berlin,
Germany
}
\offprints{G.\ Weigelt}
\mail{weigelt@mpifr-bonn.mpg.de}
\date{Received 18 December 1997 / Accepted 10 March 1998}
\titlerunning{Speckle--masking observations of IRC\,+10\,216}
\authorrunning{Weigelt et al.}
\maketitle
\begin{abstract}
We present the first K$^{\prime}$-band image of the carbon star 
IRC\,+10\,216 with
76\,mas resolution.  The diffraction-limited image was reconstructed
from 6\,m telescope speckle data using the speckle masking bispectrum
method.  The image shows that the dust shell of IRC\,+10\,216 is
extremely clumpy.  Five individual clouds within a $0\farcs21$ radius
of the central star have been resolved for the first time.
On the basis of consistent theoretical models we argue that these
structures are produced by circumstellar dust formation.
The fragmentation of the shell structure gives most likely
direct evidence for an inhomogeneous mass-loss process which may be
interpreted in terms of large-scale surface convection-cells
(Schwarzschild \cite{Schwschil_75})
being a common phenomenon for red giants.
\end{abstract}
\section{Introduction}
IRC\,+10\,216 (CW Leo) is the nearest and best--studied carbon star.  It
is a long--period variable star with a period of about 650\,d and a
spectral type of C9,5 (see e.g.\ Olofsson et al.\ \cite{OlofJoHj_82}). 
Estimates of its distance range from 100\,pc (Zuckerman et al.\
\cite{ZuckDyCl86}) to 290\,pc (Herbig \& Zappala \cite{HerbZa70}). 
IRC\,+10\,216 is surrounded by a dust shell which is
expanding at $v_{\rm exp} \approx 15\,$kms${^{-1}}$,
thereby carrying a mass loss rate of
$\dot{M} \approx 2-5\,10^{-5}M_{\odot}\,$yr$^{-1}$ (e.g.\ Loup et~al.\
\cite{LoupFoOm_93}).  The first high-resolution IR observations of the
dust shell of IRC\,+10\,216 were reported by
Toombs et al. (\cite{ToomBeFr_72}), 
McCarthy et al.\ (\cite{McCaHoLo80}, \cite{McCaMcBa90}),
Mariotti et al. (\cite{MariChSi_83}), 
Dyck et al.\ (\cite{DyckZuLe_84}, \cite{DyckHoZu_87}, \cite{DyckBeHo_91}),  
Ridgway \& Keady (\cite{RidgKe88}),  
Christou et al.\ (\cite{ChriRiBu_90}),  
Le~Bertre et al. (\cite{LeBeMaRe88}), 
Danchi et al. (\cite{DancBeDe_94}),  
Osterbart et al. (\cite{OsteBaWe_97}), and 
Weigelt et al. (\cite{WeigBaHo_97}).
Detailed radiation transfer calculations for IRC\,+10\,216 have been
carried out, for instance, by Groenewegen (\cite{Groe97}) using
a large amount of spectroscopic and visibility data.
Consistent time--dependent models describing the circumstellar shells
of dust forming long--period variables have been presented by
Fleischer et~al.\ (\cite{FleiGaSe92}).  A general result of these
models is the formation of discrete dust layers, causing pronounced
time--varying, step--like surface intensity distributions
(Winters et~al.\ \cite{WintFlGa_95}).

In this {\it Letter} we present diffraction-limited 76\,mas speckle
masking observations of the clumpy dust shell of IRC\,+10\,216.  We
speculate about the origin of these structures in the light of
consistent time--dependent model calculations and discuss the red
giants' large-scale surface convection (Schwarzschild
\cite{Schwschil_75}) as a possible mechanism for inhomogeneous mass
loss.
\section{Observations and speckle masking results}
\begin{figure}
  \setlength{\unitlength}{1mm}
  \begin{picture}(82,164)(0,0)
    \put(0,82){ \resizebox{85mm}{!}{\includegraphics{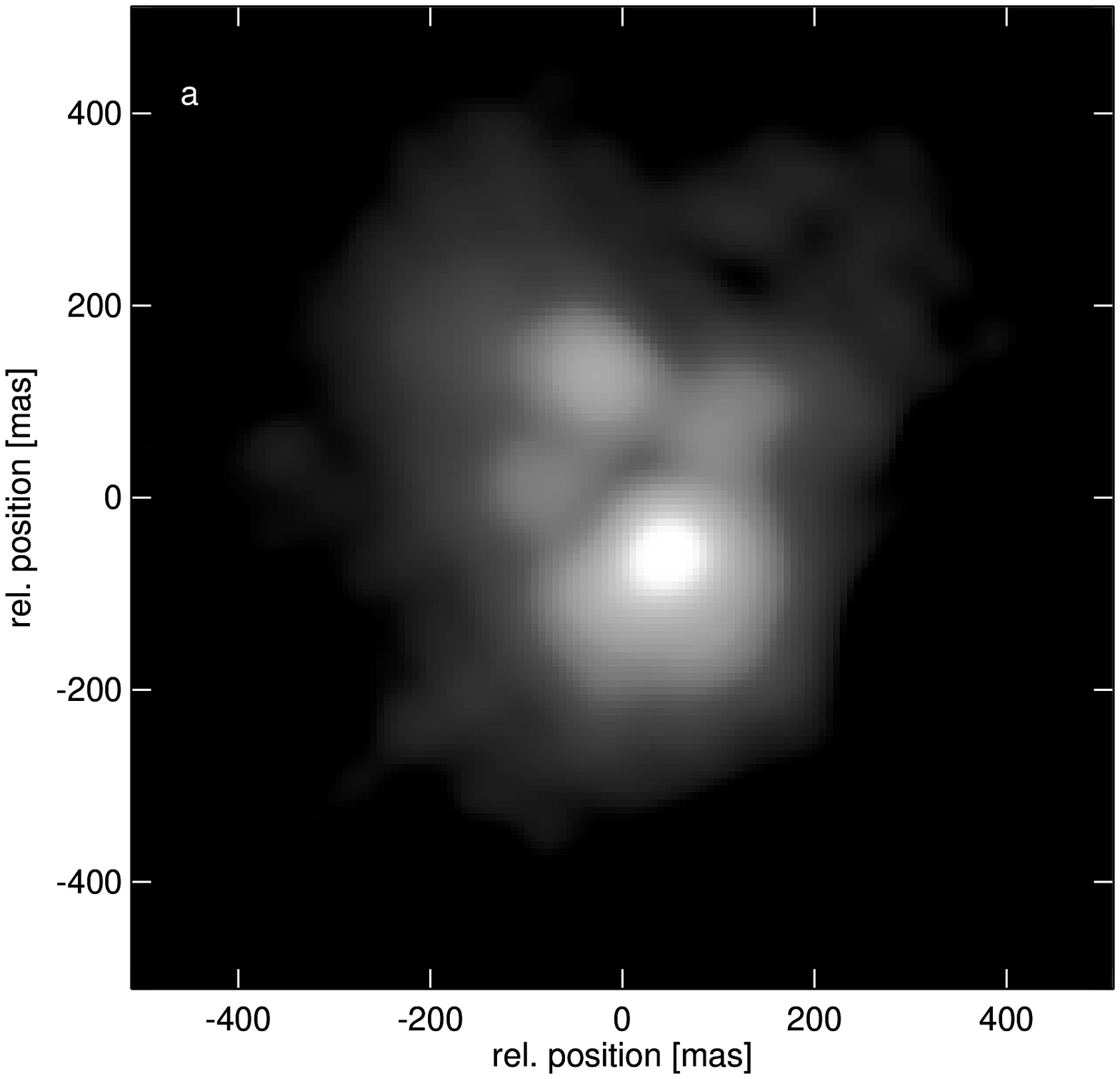} } }
    \put(0,0){ \resizebox{85mm}{!}{\includegraphics{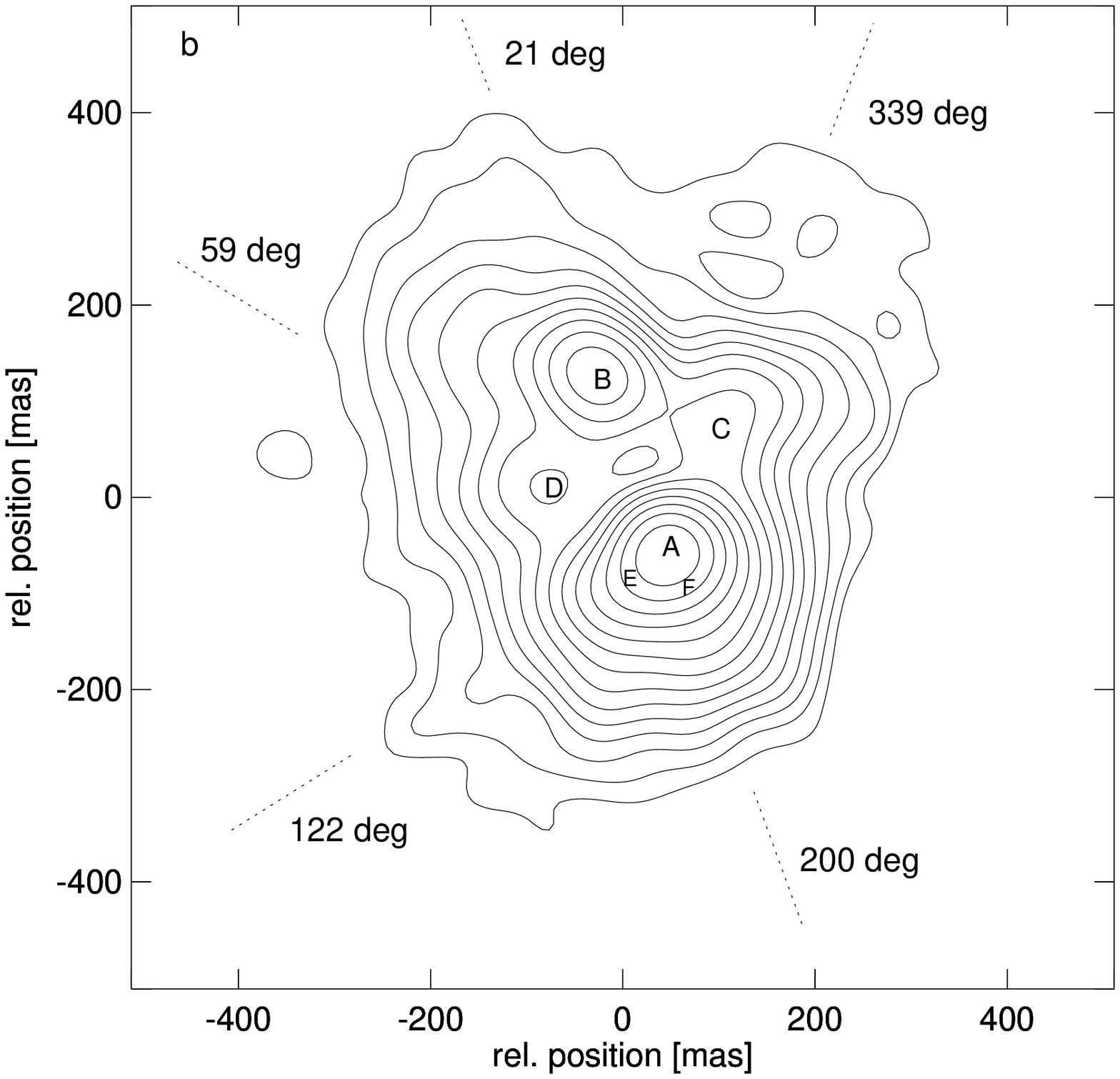} } }
  \end{picture}
\protect\vspace*{-4mm}
  \caption{
    Speckle masking observation of IRC\,+10\,216:  
    {\bf a} K$^{\prime}$-band speckle masking reconstruction, 
    {\bf b} contour plot of {\bf a} with the denotation of the
    components (the contour levels are plotted in steps of $0.3^{\rm m}$). 
    The dotted lines indicate the cut directions taken for
    Figs.~\protect\ref{imacuts}a--e
    \label{recon}
  }
\protect\vspace*{-4mm}
\end{figure}
\begin{figure}
  \setlength{\unitlength}{100mm}
  \begin{picture}(1,1)(0,0)
  \put(0,0){ \resizebox{!}{100mm}{\includegraphics{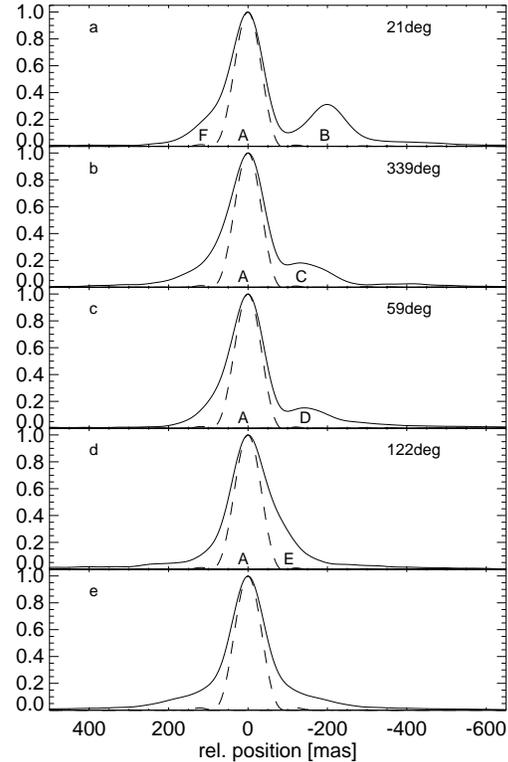} } }
  \end{picture}
  \caption{
    Intensity cuts (solid lines) through the image in
    Fig.~\protect\ref{recon} in the indicated directions.  The cut
    directions are shown by dotted lines in Fig.~\protect\ref{recon}b. 
    Dashed curves are cuts through the point spread function
    reconstructed from two different data sets of the reference star. 
    In plot e the azimuthal averages of IRC\,+10\,216 and the PSF are
    shown
    \label{imacuts}
  }
\end{figure}
The IRC\,+10\,216 speckle interferograms were obtained with the 6\,m
telescope at the Special Astrophysical Observatory on April 3,
1996 (variability phase 0.10).  The speckle interferograms were
recorded with our
NICMOS 3 camera through a standard K$^{\prime}$
filter with a center wavelength of 2.17\,$\mu$m and
a bandwidth of 0.33\,$\mu$m.  968 IRC\,+10\,216 speckle interferograms
and 1496 reference star interferograms (HIC\,51133) 
were recorded.  The observational parameters were as follows:
  exposure time per frame 70\,ms,    
  pixel size 14.6\,mas, and
  seeing (K$^\prime$) $\sim$\,2\farcs5.
Fig.~\ref{recon} shows the diffraction-limited K$^{\prime}$ 
image of IRC\,+10\,216 which
was reconstructed from the speckle
interferograms using the speckle masking bispectrum method
(Weigelt \cite{Weig77},
Lohmann et al.\ \cite{LohmWeWi83}, 
Weigelt \cite{Weig91};
note that the image published by Weigelt et al.  (\cite{WeigBaHo_97})
was actually a K$^{\prime}$ image, not an H-band image as was
incorrectly printed).  In addition to the dominant central object A,
three bright knots (called B, C, and D) can be identified.  The
separation of B, C, and D from the central object are
$\sim\,203$\,mas, 137\,mas, and 136\,mas, respectively.  The position
angles of B, C, and D are 21\degr{}, 339\degr{}, and 59\degr{},
respectively.  The reconstructed image shows that the dust shell of
IRC\,+10\,216 is extremely clumpy.  The asymmetry of the dominant
central object suggests that there are at least two
additional clouds E and F (not fully resolved) at separations much
smaller than the separations of B, C, and D (i.e., at separation
$<$\,100\,mas).  The position angles of E and F are $\sim\,122$\degr{}
and 200\degr{}, respectively.  The objects A to F are located inside a
larger nebulosity which looks like a bipolar, X--shaped nebula with an
approximately NS polar axis (or position angle 10 to 20\degr{}, in
agreement with Kastner \& Weintraub \cite{KasWein_94}).  At the 2\%
intensity level the nebulosity extends over $0\farcs8$ in NS and
$0\farcs6$ in EW direction.  The considerable brightness of cloud B
($\sim5$\% of the total flux) can, for instance, be explained if we
assume that the central star is obscured by a dust disk seen nearly
edge-on.  The smallest radius of the dominant
central object (A+E+F) was measured for position angle 21\degr{}
(see Fig.~\ref{imacuts}a).  At position angle
21\degr{} the radius of the central object was determined to be
approximately 25\,mas$\pm$8\,mas which is an upper limit for the
radius of the central star, as the observed central object A could
consist of both the stellar disk plus nearby or foreground dust
clouds.
\section{Discussion}
\subsection{Evolutionary status}
IRC\,+10\,216 is in a very advanced stage of its AGB evolution due to
its low effective temperature, long period and high mass-loss rates. 
Its carbon-rich chemistry indicates that a significant number of
thermal pulses with corresponding dredge-up events did take place. 
The mass of the hydrogen-exhausted core, $M_{\rm H}$, can be expected
to be already close to the later final mass, due to the high mass-loss
rates and limited core growth-rates per pulse of only a few $10^{-3}
M_{\odot}$ in its stage of evolution (depending on mass)
partially compensated or even canceled by dredge-up episodes.  The
conjecture that IRC\,+10\,216 has entered a phase immediately before
moving off the AGB seems to be supported by its non-spherical
appearance (Fig.~\ref{recon}; see also Kastner \& Weintraub
\cite{KasWein_94}).  In contrast to their progenitors, AGB successors
often expose prominent features of asphericities, mostly in
axisymmetric geometry.  Note, that IRC\,+10\,216 is already
considerably elongated in NS direction probably even with a
{\it bipolar} structure (Fig.~\ref{recon}).

The measured bolometric flux $S$ at maximum light
($S$\,$=$\,
2.1\,$\cdot$\,$10^{-8}$\,${\rm Wm}^{-2}$;
Sopka et al.\ \cite{SopHilJaf_85})
leads to $11000 < L/L_{\odot} < 18900$  for recent distance 
estimates of $130\,{\rm pc} < d < 170\,{\rm pc}$
(Le~Bertre \cite{LeBertre97},
Winters et~al.\ \cite{WintDoSe94}).
Introducing these luminosities into the core-mass luminosity relation will 
give upper limits for $M_{\rm H}$ since $S$ requires 
corrections for the mean variability phase and the thermal-pulse cycle phase.
Evolutionary models of Bl\"ocker (\cite{Bl_95}) give
$M_{\rm H}$\,$=$\,$0.65 M_{\odot}$ for $L$\,$=$\,$10000 L_{\odot}$. 
Note, that the $M_{\rm H}$-$L$ relation
breaks down for massive AGB models (Bl\"ocker \& Sch\"onberner \cite{BlScho_91})
due to the penetration of the envelope convection into the hydrogen-burning 
shell (``hot bottom burning'', HBB). Accordingly,
the upper luminosity value indicates  $M_{\rm H} \la 0.8M_{\odot}$
and possibly HBB.
Since the present core mass will not deviate much from the final mass, 
it can be applied in initial-final mass relations.
Taking the AGB calculations of Bl\"ocker (\cite{Bl_95}) and 
Vassiliadis \& Wood (\cite{VassWoo_93}),
resp., we finally arrive at initial masses  lower than
$2.5 - 3\,M_{\odot}$ for $d = 130$\,pc
and $4 - 4.5\,M_{\odot}$ for $d = 170$\,pc.
The chemistry of the circumstellar envelope gives further constraints. 
Guelin et al.\ (\cite{GueForVal_95}) 
compared the observed isotopic abundance ratios 
with evolutionary models. Particularly 
the C, N and O isotopic ratios led to the conclusion that the initial mass
ranges between 3 and $5 M_{\odot}$, and that moderate HBB has taken
place, favouring an initial mass close to $4\,M_{\odot}$.
\subsection{Discrete dust layers}
\begin{figure}
  \resizebox{80mm}{!}{\includegraphics{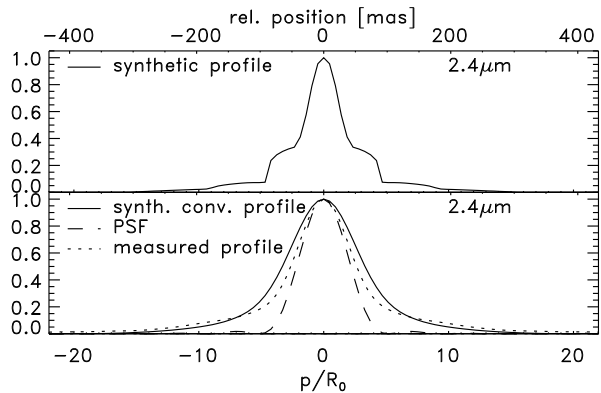}}
\caption[]{
	   Synthetic intensity profile at $\lambda=2.4\mu$m (upper
	   panel) resulting from a model calculation (see text).  The
	   lower panel shows the convolution (solid line) of the
	   synthetic profile with the PSF of the 6\,m telescope, the
	   PSF of the 6\,m telescope (dashed line), and the azimuthal
	   average of the measured intensity (dotted line)}
\label{intpic}
\end{figure}

As a demonstrative example Fig.~\ref{intpic} shows a one--dimensional
synthetic intensity profile resulting from a consistent
time--dependent model calculation for a carbon--rich circumstellar
dust shell.  These model calculations assume spherical symmetry and
include a consistent treatment of time--dependent hydrodynamics,
chemistry, dust formation, growth and evaporation and of the radiative
transfer problem
(Fleischer et~al.\ \cite{FleiGaSe92}, Winters et~al.\
\cite{WintFlGa_94}).
A general result of the calculations is the
formation of discrete dust layers with the characteristic
step--like intensity profile shown in
Fig.~\ref{intpic} (top).  The location and height of the steps vary in
time since the dust layers are moving outwards and, thereby, become
geometrically diluted (see Winters et~al.\ \cite{WintFlGa_95}).  
In the calculation, these structures are
produced by thermal dust emission which, via the dust opacity, depends
on wavelength.  Note also, that the steps are separated by only a few
stellar radii (lower abscissa; the upper abscissa gives the
angular extension assuming $d=170$\,pc).  The
bottom diagram of Fig.~\ref{intpic} shows the intensity profile 
convolved with the ideal point spread function (PSF) of the 6\,m
telescope (FWHM diameter 76\,mas). At this resolution the 
step-like structures in the 
intensity profile disappear completely.  Comparing this
intensity profile (Fig.~\ref{intpic}, bottom) with our measured one
(same figure) shows that there is a good agreement, but the
wings of the measured profile are slightly higher. 

The most striking structures in our image (Fig.~\ref{recon})
are the three knots B, C, and D.  Assuming a typical
stellar radius of the central source of $R_{\ast} = 5 \cdot 10^{13}$cm
and $d = 170$\,pc (Winters et~al.\ \cite{WintDoSe94}), the
(tangential) separation of the knots from the central peak is
10\,$R_{\ast}$ (B) and 7\,$R_{\ast}$ (C,D).  In terms of the models,
this could be interpreted as knots B, C, and D being connected to an
outer dust layer, whereas knots E and F belong to the next layer
inwards.  This interpretation requires the fragmentation of
inhomogeneous dust layers or that the knots result from spatially
bounded separate dust formation events.  Since dust nucleation is
extremely sensitive to the local kinetic gas temperature, dust
formation could be caused locally even by small temperature fluctuations.
The radial velocity of the expanding dust shell is approximately 
15\,km\,s$^{-1}$. 
This corresponds to $\sim$\,3\,AU/year or 18\,mas/year at a distance of 170\,pc
for a movement perpendicular to the line--of--sight.  Thus, if
connected to this expansion, knot B should have formed at least
11.6\,yr ago, while C and D would be $\geq\,7\,$yr old.  In terms of
the pulsation period ($P\,\sim\,650$\,d) this would 
correspond to a time 
scale for the formation of new dust layers (or knots) of
$\Delta\,t\,\geq\,2.5\,P$.  Then, the structures E and F would have been
formed $\sim\,1.5\,P$ ago.  The formation of new dust layers on time
scales longer than the pulsation period is a common
phenomenon of the model calculations (e.g.\ Fleischer et~al.\
\cite{FleiGaSe92}, Winters et~al.\ \cite{WintFlGa_94},
\cite{WintFlGa_95}).  Therefore, 
future observations can be used to test
this model and to determine the dust formation frequency and the
tangential velocity of the structures. 
\subsection{Inhomogeneuos mass loss}
Since the present observations reveal that IRC\,+10\,216's
shell structure is {\it highly fragmented} in the immediate
stellar vicinity, there seems to be evidence for an already {\it
inhomogeneous mass-loss process}.  Inhomogeneously outflowing matter
implies corresponding stellar surface inhomogeneities which may be
caused by magnetic activity, global pulsations or large-scale
photospheric convection.  In particular, the latter seems to be a
common phenomenon of far-evolved stars.

Schwarzschild (\cite{Schwschil_75}) 
showed that the typical horizontal size of a 
solar granule, $x_{\rm gran}$, is given by  characteristic depth scales 
of the layers below the photosphere. 
With the pressure scale height, $H_{\rm p}$,
as the major depth scale and assuming 
that the ratio of $x_{\rm gran}/H_{\rm p}$ is constant he found that
for red giants the dominant convective elements
become so large that only a few of them can occupy the surface at any 
time leading to large temperature variations on the surface and 
concomitant brightness fluctuations. Due to the prominent temperature contrasts
at the surface the emitted radiation is highly anisotropic leading to a
polarization of the light scattered by circmstellar dust. 
Schwarzschild (\cite{Schwschil_75}) already supposed 
that mass ejection is triggered by photospheric convection and Dyck et al.\
(\cite{DyckHoZu_87}) outlined its possible importance for IRC\,+10\,216.

Indeed, based on a linear stability analysis of convective modes in the envelopes of 
red giants, Antia et al.\ (\cite{AntChitNar_84}) 
found the pressure scale height to be the prevailing depth
scale leading to dominant convective elements which are of 
comparable size to the stellar radius. 
More recently, Freytag et al.\ (\cite{FreyHolStef_97})
presented models for convection zones 
of main-sequence stars and subgiants 
with spectral type F to K 
based on 2D numerical
radiation hydrodynamics calculations. 
They found a tight correlation between the characteristic
photospheric scale height $H_{{\rm p}0} = R_{\rm gas} T_{\rm eff}/g$ 
and the size of the granules, $x_{\rm gran}$, viz.\ 
$x_{\rm gran} \approx 10 H_{{\rm p}0}$
covering more than two orders of magnitudes in gravity. Due to this robustness 
a (cautious) extrapolation to the red giant regime seems to be justified. 
For IRC\,+10\,216 (with $T_{\rm eff} = 2200$\,K,
$R = 700 R_{\odot}$, and $M = 0.8 M_{\odot}$) this leads to a typical 
granule size of $x_{\rm gran} \approx 0.82 R_{\ast}$ allowing the whole 
surface to be occupied by, at most, a few granules. 

The  resulting temperature fluctuations can be expected to be in the range 
of  up to several hundred Kelvin (Antia et al.\ \cite{AntChitNar_84})
being large enough to cause observable 
brightness fluctuations and to influence  the formation of the 
shock- and dust-driven stellar wind and, therewith, the shape of the 
circumstellar shell. 
Thus, one further implication 
of large-scale surface inhomogeneities could be a corresponding large-scale
fragmentation of the outflowing matter possibly leading to knots within 
the multiple shell structure as observed for IRC\,+10\,216. 
\section{Conclusion}
Our speckle masking reconstruction of IRC\,+10\,216 shows 
that the circumstellar dust shell of IRC\,+10\,216
consists of at least five individual dust clouds B to F within a
$0\farcs21$ radius of the central star.  With $d=170$\,pc and
$R_{\ast} = 5 \cdot 10^{13}$cm, the tangential separations of the
clouds B, C, and D from the central star correspond to
$\sim$\,34\,AU $\sim 10\,R_{\ast}$ (B) and 
23\,AU $\sim 7\,R_{\ast}$ (C,D).  Time-dependent calculations for
circumstellar envelopes show the formation of multiple-shell
structures, and the synthetic intensity
profiles agree well with the measured ones. 
The prominent clumpiness of these very inner shells 
gives evidence for an already inhomogeneous mass loss which may be
intimately linked with large-scale surface inhomogeneities
possibly induced by giant surface convection-cells.
\begin{acknowledgements}
We thank B.\ Freytag for valuable discussions on surface convection zones.
J.M.W.  and A.J.F.  thank the Konrad--Zuse--Zentrum f\"{u}r
Informationstechnik Berlin and the HLRZ, J\"ulich for a generous grant
of computing time.  A.J.F.\  was supported by the DFG (grant
Se~420/8--1) and J.M.W.\  by the BMBF (grant 05 3BT13A 6). 
\end{acknowledgements}
\end{document}